\documentclass[fleqn,twoside]{article}
\usepackage{espcrc2}

\usepackage{graphicx}
\def\rmO{{\rm O}}
\def\za{Z_{\rm A}}
\def\R{\mathcal R}
\def\rmd{{\rm d}}

\def\op{{\cal O}}

\def\ba{b_{\rm A}}

\newcommand{\AmS}{{\protect\the\textfont2
  A\kern-.1667em\lower.5ex\hbox{M}\kern-.125emS}}

\title{
{
\vspace{-4.0cm} \normalsize \hfill
\parbox{30mm}{HU-EP-03/63\\DESY 03-138\\SFB/CPP-03-34\\[3mm] September 2003}
}\\[15mm]
Non--perturbative renormalization of the axial current\\
       with improved Wilson quarks
       \thanks{Talk presented by R. Hoffmann at Lattice 2003}}

\author{R. Hoffmann\address[HU]{Institut f\"ur Physik, Humboldt Universit\"at,\\ 
        Newtonstr. 15, 12489 Berlin, Germany},
        F. Knechtli\addressmark,
        J. Rolf\addressmark,
        R. Sommer\address{DESY, Platanenallee 5, 15738 Zeuthen, Germany}
        and        
        U. Wolff\addressmark[HU]}
       
\begin{document}
\thispagestyle{empty}
\begin{abstract}
\vspace{-0.4pc}
We present a new normalization condition for the axial current,
which is derived from the PCAC relation with non--vanishing
mass. Using this condition reduces the $\rmO(r_0 m)$ corrections
to the axial current normalization constant $\za$ for an easier chiral
extrapolation in the cases, where simulations at zero quark--mass
are not possible. The method described here also serves as a
preparation for a determination of $\za$ in the full two--flavor theory.
\vspace{-1.1pc}
\end{abstract}

\maketitle

\section{Introduction}

\vspace*{-2mm}

Due to the explicit breaking of chiral symmetry in a lattice
theory with Wilson quarks a finite renormalization
constant $\za$ for the isovector axial current is needed so that
Ward identities take their canonical form up to small lattice
corrections. Otherwise a safe extrapolation to the continuum
can not be achieved for physical quantities involving this current.

The function $\za(g_0)$ \cite{Luscher:1996jn} currently used by the
ALPHA collaboration was obtained by requiring that certain chiral Ward
identities are satisfied in the $\rmO(a)$ on--shell improved lattice
theory up to $\rmO(a^2)$ cutoff effects. This normalization condition
is evaluated at zero quark mass and hence the mass term in the
PCAC relation was dropped.

However, even in the framework of the Schr\"odinger functional simulations
at vanishing quark mass are not possible if the physical volume, or equivalently the renormalized coupling, is too large. For the large couplings one therefore evaluates the
normalization condition at finite quark mass and extrapolates to
the chiral point. Since the mass term was ignored in the derivation
of the normalization condition this introduces an error of
$\rmO(r_0m)$ in $\za$. The Sommer scale $r_0$~\cite{Sommer:1993ce}
is used as a typical hadronic scale.

Here we present a normalization condition for the axial
current, which is derived using the full PCAC relation.
This reduces the $\rmO(r_0m)$ errors in $\za$ to $\rmO(am)$ and
therefore results in a much flatter chiral extrapolation.\\[-5mm]

\section{The axial Ward identity}

\vspace*{-2mm}

By performing local infinitesimal symmetry
transformations of the quark and anti--quark fields in the Euclidean
functional integral one derives the Ward identities associated with
the chiral symmetry of the continuum action. An axial transformation of the
isospin doublet quark fields gives the partially conserved axial
current (PCAC) relation
\begin{equation}
\ \langle\partial_\mu A_\mu^a(x)\op\rangle-2m\langle P^a(x)\op\rangle
=-\langle\delta\op\rangle
\label{PCAC},
\end{equation}
where $\delta\op$ is the variation of some operator $\op$.
$A_\mu^a(x)$ and $P^a(x)$ are the isovector axial current and isovector
pseudoscalar density, respectively.

Let $\R$ be the space--time region where the chiral transformation is
applied. The axial current $A_\nu^b(y)$ is inserted as an internal operator
($y\!\in\!\R$). If $\op_{\rm ext}$ denotes a polynomial in the
basic fields outside this region the integrated form of the axial
current Ward identity is
{\small
\begin{eqnarray}
\int_{\partial \R}\!\!\!\textstyle\rmd\sigma_\mu(x)\,\Big\langle
A_\mu^a(x)A_\nu^b(y)\op_{\rm ext}
\Big\rangle\qquad\qquad\nonumber\\
\qquad\qquad-2m\int_R\!\!\rmd^4x\, \Big\langle
P^a(x)A_\nu^b(y)\op_{\rm ext}\Big\rangle\nonumber\\
\qquad\qquad\qquad\qquad=i\epsilon^{abd}\Big\langle V_\nu^d(y)\op_{\rm
ext}\Big\rangle.\label{WI}
\end{eqnarray}
}
The right--hand side of (\ref{WI}) originates from the variation of
the internal operator $A_\nu^b(y)$ since under chiral
transformation the axial current rotates into the vector current
$V_\mu^a(x)$.

\section{Normalization conditions on the lattice}

A normalization condition for the axial current on the lattice is
derived by demanding that eq.~(\ref{WI}) in terms of the renormalized
currents is still valid up to lattice artifacts for a given choice of
$\R$ and $\op_{\rm ext}$.

We use Schr\"odinger functional boundary conditions in
$\rmO(a)$-improved lattice QCD \cite{Luscher:1992an,Sint:1993un}, i.e.
we have an $L^3\times T$ space--time cylinder with periodic boundary
conditions in the spatial directions and fixed (Dirichlet) boundary
condition in the temporal direction.

We choose $\R=L^3\times[T/3,2T/3]$ and an $\op_{\rm ext}$, which creates
pseudo-scalar states at the upper and lower temporal boundary.
In the renormalization scheme 
we employ \cite{Luscher:1996sc,Jansen:1995ck} the normalization
condition now gives a relation between a set of 
(improved) correlation functions on the lattice and the renormalization
factor $\za(1+\ba am_q)$.

If the mass is set to zero in
(\ref{WI}) with our choice for $\op_{\rm ext}$ the lattice
version of equation (\ref{WI}) becomes
\begin{eqnarray}
\ \za^2f_{AA}^I=f_1,\label{normcon1}
\end{eqnarray}
where $f_1$ is the correlator connecting the pseudo-scalar 
boundary states and $f_{AA}^I$ is the correlation function corresponding
to the surface integral in (\ref{WI}) in terms of the improved
lattice currents.

If the mass term in the PCAC relation (\ref{PCAC}) is not neglected
one has to evaluate an additional correlation function $\tilde f_{PA}^I$
corresponding to the second term in (\ref{WI}). For this correlator
on--shell $\rmO(a)$ improvement breaks down due to contact terms.
The normalization condition thus obtained is
\begin{eqnarray}
\ \za^2(1\!+\!\ba am_q)^2\Big(f_{AA}^I
-2m\tilde f_{PA}^I\Big)=f_1.\label{normcon2}
\end{eqnarray}
In the improved theory the normalization conditions (\ref{normcon1})
and (\ref{normcon2}) result in an intrinsic uncertainty of
$\rmO(a^2)$ in $\za$. Condition (\ref{normcon1}) will lead to an
additional error of $\rmO(r_0m)$ when evaluated at finite quark mass
whereas (\ref{normcon2}) will lead to $\rmO(am)$ errors from the 
contact terms in the correlator multiplying the quark mass and also
from $\ba$, which we only know perturbatively.

\section{Testing the new normalization condition}

We choose the Schr\"odinger functional setup with $T=\frac94L$,
$\theta=0$ and vanishing background field $(C=C'=0)$ and
evaluate the normalization conditions on $8^3\times18$
quenched lattices with several different values of the hopping
parameter.

\begin{figure}[!ht]
\begin{center}
\vspace*{-8mm}
\hspace*{-1mm}\includegraphics[width=80mm]{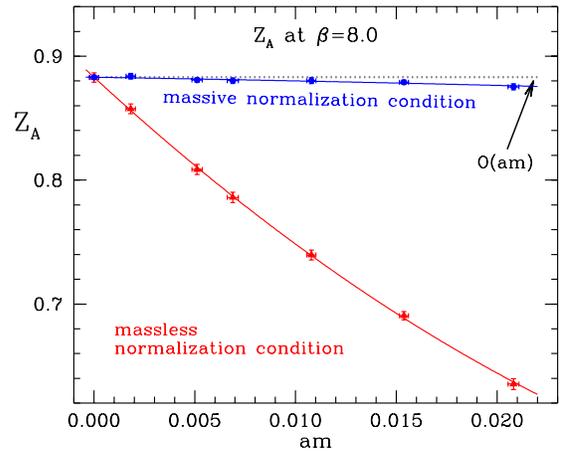}
\vspace*{-12mm}
\caption{Comparison of old (massless) and new (massive)
normalization condition at small coupling. The dotted line is the
result for $\za$.\label{fig:beta8}}
\vspace*{-8mm}
\end{center}
\end{figure}

Figure \ref{fig:beta8} shows the results from both normalization
conditions as a function of the PCAC mass for $\beta=8.0$ on the
same set of lattices. In this case a simulation at the chiral point
is possible in the Schr\"odinger functional. As expected the quark
mass dependence is strongly reduced when the new normalization
condition is used. Also the statistical error in $\za$ is much smaller
with the massive condition.

\begin{figure}[!ht]
\begin{center}
\vspace*{-0mm}
\hspace*{-1mm}\includegraphics[width=74mm]{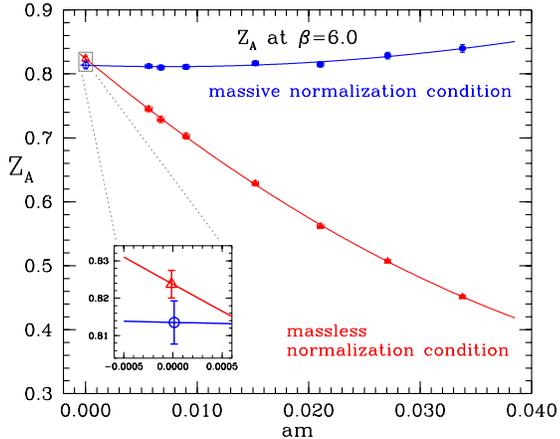}
\vspace*{-9mm}
\caption{Both normalization conditions evaluated at $\beta=6.0$.
The extrapolated values are shown in the insert.\label{fig:beta6}}
\vspace*{-13mm}
\end{center}
\end{figure}

More interesting is the case of $\beta=6.0$, where $\za$ has to be
extracted from a chiral extrapolation using a quadratic fit.
Figure \ref{fig:beta6} shows
that in this case the curvature is more pronounced but the mass
dependence for the new method is still quite flat. The error in
the extrapolated value for $\za$ is however larger for the massive
normalization condition:\\

\begin{equation}
\begin{tabular}{ll}
massive  & $\,  \za=0.8135(58)$  \\[0mm]
massless & $\,  \za=0.8237(37)$
\end{tabular}\label{full}
\end{equation}

\pagebreak

\section{Disconnected quark diagrams}
On closer inspection the reason for this larger error can be found
by looking at the possible Wick contractions for the correlation
functions $f_{AA}^I$ and $\tilde f_{PA}^I$. The structure
of the external operator used leads to disconnected quark diagrams
where the two current insertions each couple to one temporal end of the
lattice only.

Those cause large statistical fluctuations especially in the volume
correlator $\tilde f_{PA}^I$, which make the extraction of $\za$
from the new normalization condition statistically more difficult.

These diagrams can be suppressed by using a different external operator
in the derivation of the normalization conditions. This operator requires
the existence of a third quark species (spectator) and can therefore
be used only as an approximation in a theory with $N_f<3$. We are
working on an extension of this argument.

\begin{figure}[!ht]
\begin{center}
\vspace*{-6mm}
\hspace*{-1mm}\includegraphics[width=74mm]{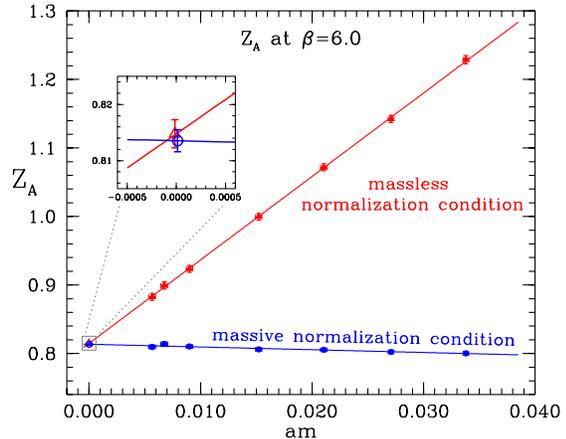}
\vspace*{-9mm}
\caption{Both normalization conditions using only the connected diagrams.
\label{fig:beta6con}}
\vspace*{-13.8mm}
\end{center}
\end{figure}

Figure \ref{fig:beta6con} shows the result from both conditions if only
connected quark diagrams are used. Note that in this case a linear extrapolation
in the quark mass is possible. The extrapolated values for $\za$ in this
case are\\[-4mm]
\begin{equation}
\begin{tabular}{ll}
massive  & $\,  \za=0.8135(20)$  \\[0mm]
massless & $\,  \za=0.8148(25)$. \\[-2mm]
\end{tabular}
\end{equation}

Here the new condition not only results in a flat extrapolation,
it also gives an extrapolated value of $\za$, which is consistent
with (\ref{full}) and has a smaller statistical error.

\vspace*{-0.5mm}

\section{Conclusions}

The new normalization condition including the mass
term has only a very small dependence on the quark mass.
If the spectator quark argument can be 
extended to the two flavor case it is therefore a suitable tool
for calculating the axial current normalization constant in the cases
were simulations at zero quark mass are not possible.

This work is also a preparation for an unquenched
determination of $\za$.
Due to algorithmic problems at large values of the bare gauge coupling
and small quark masses we will benefit greatly from a reliable chiral
extrapolation.

We thank M. L\"uscher and S. Sint for valuable contributions. This work
was supported by the DFG (SFB/TR 09) and the GK271.


\begin{thebibliography}{9}
\bibitem{Luscher:1996jn}
M.~L\"uscher, S.~Sint, R.~Sommer and H.~Wittig,
Nucl.\ Phys.\ B {\bf 491} (1997) 344

\bibitem{Sommer:1993ce}
R.~Sommer,
Nucl.\ Phys.\ B {\bf 411} (1994) 839

\bibitem{Luscher:1992an}
M.~L\"uscher, R.~Narayanan, P.~Weisz and U.~Wolff,
Nucl.\ Phys.\ B {\bf 384} (1992) 168.


\bibitem{Sint:1993un}
S.~Sint,
Nucl.\ Phys.\ B {\bf 421} (1994) 135.

\bibitem{Luscher:1996sc}
M.~L\"uscher, S.~Sint, R.~Sommer and P.~Weisz,
Nucl.\ Phys.\ B {\bf 478} (1996) 365.

\bibitem{Jansen:1995ck}
K.~Jansen {\it et al.},
Phys.\ Lett.\ B {\bf 372} (1996) 275
\end{thebibliography}
\end{document}